\documentclass[smallextended]{svjour3}       

\usepackage{bm}

\smartqed  
\usepackage{graphicx}
\def\dd{{\rm d}}
\def\DD{{\rm D}}
\def\bc{\begin{center}}
\def\ec{\end{center}}
\newcommand{\be}[1]{\begin{equation}\label{#1}}
\newcommand{\ee}{\end{equation}}
\def\ben{\begin{displaymath}}
\def\een{\end{displaymath}}
\def\ban{\begin{eqnarray*}}
\def\ean{\end{eqnarray*}}
\def\TT{\mbox{\boldmath$T$\unboldmath}}
\def\GG{\mbox{\boldmath$G$\unboldmath}}
\def\PP{\mbox{\boldmath$P$\unboldmath}}
\newcommand{\weglassen}[1]{}

\newcommand{\partq}[2]{\frac{\partial #1}{\partial #2}}
\newcommand{\Gam}[2]{{\Gamma^#1}_{#2}}
\newcommand{\Gef}[4]{C^#1{}_#2{}^#3{}_#4}
\def\detR{{\rm - det}\ R_{..}}
\def\sqdetR{\sqrt{{\rm - det}\ R_{ab}\mbox{~}}}
\def\sqdetg{\sqrt{{\rm - det}\ g_{ab}\mbox{~}}}
\begin{document}
\title{On constructing purely affine theories with matter}
\author{Jorge L. Cervantes--Cota \and D.--E. Liebscher}
\institute{Jorge L. Cervantes--Cota \at 
Depto.de F\'isica,
Instituto Nacional de Investigaciones Nucleares,\\
Apartado Postal 18--1027, 11801 CDMX, M\'exico
\and
D.--E. Liebscher  \at 
Astrophysikalisches Institut Potsdam,
An der Sternwarte 16, D-14482 Potsdam}
\maketitle

\begin{abstract}
We explore ways to obtain \textit{the very existence} of a space--time 
metric from an action principle that does not refer to it a priori. Although 
there are reasons to believe that only a non--local theory can viably achieve 
this goal, we investigate here local theories that start with Schr\"odinger's
purely affine theory \cite{SER50}, where he gave reasons to set
the metric proportional to the Ricci curvature aposteriori. When we leave the 
context of unified field theory, and we couple the non--gravitational matter 
using some weak equivalence principle, we can show that the propagation of 
shock waves does not define a lightcone when the purely affine theory is 
local and avoids the explicit use of the Ricci tensor in realizing the weak 
equivalence principle. When the Ricci tensor is substituted for the metric, the 
equations seem to have only a very limited set of solutions. This backs the 
conviction that viable purely affine theories have to be non--local.

\keywords{Affine theories \and local gravity theories.}

\end{abstract}


\section{Introduction}

Purely affine theories became a topic in Relativity through the search for
a theory unifying gravitation and electromagnetism.  A central question of a
purely affine theory is the generation of a metric in the course of the 
evaluation of the field equations.   This property of an a posteriori generation 
of the space--time metric is a central issue for a relativistic implementation
of a Mach principle, e.g. as a Mach-type symmetry breakdown to locally Lorentz invariant 
theories \cite{LYW79,LDE81B,LDE88C}. The implementations of the Mach principle 
into a relativistic theory of gravity have found different aspects and different 
directions to explore \cite{BaPf95,THJ72}.  Considering a Mach type symmetry
breakdown to locally Lorentz invariant theories, the important aspect is that 
the light-cone is the structure that should be generated through that 
break--down. This implies that the metric structure itself should not \textit{a priori} enter in the
gravitation theory.  The metric structure, together with the existence 
of a light--cone, should be the outcome of the theory. In this context,
the distribution of matter in the surrounding universe represents the classical 
vacuum for the local neighborhood that breaks the \textit{at-least} affine invariance 
of vector spaces to the Lorentz invariance; for a local 
breakdown, see Ref. \cite{NSD88A}. Although a Mach type symmetry
breakdown should require an a priori non-local theory \cite{LYW79,Li88}, it is 
surely useful to consider the known local pre-metric theories anew.
This is the reason why we do not use the in other respect successful
path to extend the metric theory to a metric-affine theory \cite{H}, but return again to 
purely affine theories.

In our approach, the metric is only expected to be a second-order tensor that
appears in the simplest equations of motion, like the motion of pole particles or the
propagation of shock waves of any field. These equations should be compared with the 
corresponding equations of General Relativity (GR) in order to identify the light-cone structure, or the
projective structure used by Ehlers, Pirani, and Schild \cite{EhPiSc72}. In the present work we shall
consider shock waves because the appearance of the light-cone structure is our central point.
These shockwaves can be matter shocks as well as pure gravitational shocks. In any case,
we need only the simplest approximations.
Symmetry properties of the tensor that is to be identified as metric 
are a second-order problem and will not be discussed here.

The first observation is that we cannot avoid the use of a connection  
$\Gam{a}{bc}$ even  when the use of an ordinary metric is avoided: The pure 
definition of a covariant derivative requires its existence. Then, we have 
two options. First, we can formulate the problem to find a metric as some 
solution to the Weyl-Cartan problem: To find a second--order covariant 
tensor $g_{ik}$ that is covariantly constant with respect to the 
transport $\Gam{a}{bc}$ \cite{BHH97,BHH98}. The second option is to find 
independently this tensor from field equations, and consider its relation to 
the $\Gam{a}{bc}$ after\-wards. We choose this latter option here, 
because we intend to study the possibility of a dynamical definition of the metric.
Its definition through the Weyl--Cartan space problem is a priori to the 
construction of the coupling to matter and not a posteriori.
We are interested in a scheme that constructs the action without the use of the Riemannian metric so
that the latter can arise from dynamics, i.e. a posteriori.

In section \ref{secII} we construct the general type of theories that we are pursuing to deal with. In 
section \ref{secIII} the covariant field equations of our general theory are deduced. Later, in section  \ref{secIV}, we 
explore a way to find the metric tensor as a result of a schock wave. In section \ref{secV} we explore local 
actions and find the necessity to go for nonlocal actions, that are explained in section  \ref{secVI}, where also 
our final remaks are expressed.

\section{Constructing a theory} \label{secII} 

We start from a theory that defines gravitation by a connection
field.  The question is how to couple external fields and how to get the 
notion of a metric a posteriori, i.e. to find the equivalent
for the metric tensor. We expect that this a posteriori metric tensor is 
defined only to some approximation, or as a result of some symmetry breaking 
process, and that its precise definition requires particular configurations 
of the gravitational field.

Let us construct the simplest action integral for some fields $\Phi^A$, where $A$
stands for any field components without referring to their quality as 
scalar, vector, or tensor of any rank. The question of spinors in 
purely affine theory requires
particular attention, see for instance Ref. \cite{NSD88A,BTH99}. First, we look for a 
second--order field theory, i.e. for an action bilinear in the 
derivatives, ${\Phi^A}_{,k}$.  However, covariance requires the use of covariant 
derivatives, ${\Phi^A}_{;k}$, which are defined through some linear 
connection $\Gam{m}{nk}$.  The correction to the ordinary derivative for 
obtaining the covariant one is linear in the coefficients of the connection and
linear in the field,
\be{covdev}
   {\Phi^A}_{;k} = {\Phi^A}_{,k} + \Gef{A}{B}{n}{m}\ \Gam{m}{nk} \Phi^B.
\ee
where $\Gef{A}{B}{n}{m}$ are some coefficients to be determined and depend on the 
nature of the matter fields.  For instance,  when $A$ stands for indexing the 
components of a contravariant vector, then 
$\Gef{A}{B}{n}{m} = \delta^A_m\delta^n_B$. The interpretation of the
torsion and non-metricity part of the $\Gam{a}{bc}$ is a famous problem
\cite{WHE22,BSG98,BTH02}. We will consider here a general connection, not necessarily a symmetric one.

Second, the integrand must be a scalar density, so the indices of derivation have to
be compensated by some appropriate construction that provides upper indices.  In 
GR this is done through the metric tensor, more 
precisely, through its contravariant inverse, and combinations of it. Here, we have 
two options that are characterized through the use of the Ricci tensor. We shall 
consider them below.

Third, we need an invariant volume element \cite{GMM98}. When there is no metric, and 
hence no determinant of the metric tensor, the simplest choice is the determinant of 
the Ricci tensor, as A.S. Eddington pointed out in the early 1920's \cite{Eddington}. This is 
Schr\"odinger's choice too  \cite{SER46,SER47,SER48,SER48A,SER50}. With 
the simple action,
\be{sone}
S_1 = \int \sqdetR\ \dd^4 x\ ,
\ee
where $ R_{ab}$ denotes the Ricci tensor. This was used by Eddington \cite{Eddington},
but with the restriction to a symmetric connection.
Schr\"odinger obtained a theory for the general affine connection that suggested to
equate the Ricci tensor with the metric: The Ricci tensor obeys
a field equation that tells that
it is covariantly constant 
with respect to the star affinity up to the torsion of the latter. Therefore, 
Schr\"odinger postulated that 
\be{dS}
g_{ik} = \frac{1}{\lambda}R_{ik}\ .
\ee
For a unified field theory that does not explicitly contain matter
this interpretation might be satisfying, but for a theory with
explicit matter terms it is not.
Indeed, Schr\"odinger's original intention
was to get a unified field theory with no external matter at all, and the problem
was to find equivalents for the conventional matter. In the present work
we assume the gravity sector given by that of Schr\"odinger's and see how and which
ordinary matter can be coupled to such a gravitation.

We show that it is not sufficient to purely determine the Ricci tensor to act 
as metric.  The metric that is inferred by observation is that of the motion 
of matter \cite{BTH02}. This is also the lesson in particular of
all theories with more than one metric tensor \cite{LDE81B}.
Explicit matter defines an effective metric by its motion, either by the motion of
test particles or by the motion of shock waves.
We have to define test particles of the matter fields that allow
to construct an effective metric through the
Ehlers--Pirani--Schild procedure \cite{EhPiSc72}, or
we have to consider the propagation of shock fronts \cite{THJ62,PRO72}.
For the electromagnetic field, this has been stated many times 
\cite{PAS62,NWT77,JAZ79,OHF99,HOR00,HOY01}.  This is exactly our point 
of view.  We intend here to consider the relation of this 
construction --- generalized to any field --- to the Ricci tensor, that 
was Schr\"odinger's favorite choice.  It is the matter Lagrangian
that is important when we intend to define a metric.  Because it is quadratic in the
derivatives of the fields, we have to use $\sqdetR\ \dd^4x$ itself
as the invariant volume element, or alternatively we have to use fields that are
densities of weight 1/2 \cite{GMM98}.

To construct the matter part of the action within a local theory, we first recall that in 
GR this is given through
\be{stwo}
S_2 = \int L_{\rm matter}[\Phi^A, {\Phi^A}_{;k}, g^{ik}] \sqdetg \dd^4x \, .
\ee
Our construction is however performed by using $ R_{ik}$  instead of $ g_{ik}$. Then, we
have two options. First, we can try actions with Lagrangians not explicitly containing 
the Ricci tensor, where the latter enters only the volume element,
\be{stwoone}
S_3 = \int L_{\rm matter}[\Phi^A, {\Phi^A}_{;k}] \sqdetR \dd^4x  \, .
\ee
Alternatively, we may consider a matter action similar to Eq. (\ref{stwo}), but not 
necessarily implying the equality given by  Eq. (\ref{dS}),
\be{stwotwo}
S_4 = \int L_{\rm matter}[\Phi^A, {\Phi^A}_{;k}, R_{ik}] \sqdetR \dd^4x  \,\ .
\ee
Jakubiec and Kijowski \cite{JKJ89,JKJ89A} have shown that the latter action can be transformed into
GR with a different set of non-gravitational fields. This helps with respect to
the dynamical structure, but destroys the interpretation of the deliberately chosen
fields.

It is, of course, a drawback in the local action that matter has to exist \textit{locally}
in order to have a geometry defined. In the elementary vacuum, $\Phi_A \equiv 0$, the
Euler--Lagrange equation might not exist or show singular behavior.  Only the matter 
in the surrounding universe, like in Machian approaches, and not the purely
local one, should be as necessary as sufficient
to fix a geometry. However, a non--local Lagrangian will
be the next step.  First, we intend to evaluate a local action.     

\section{Covariant field equations} \label{secIII}

In this section, we explicitly show the construction of covariant field equations 
derived from the action, Eq. (\ref{stwoone}), in the case where the transformation properties
of the field components $\Phi^A$ are not yet defined.
We want to keep our formalism as
general as possible, therefore we consider our basic matter field of the following form
$\Phi^A \equiv  [{\Phi^{i_1 \cdots i_{m_1}}}_{k_1 \cdots k_{n_1}}, 
{\Phi^{i_1 \cdots i_{m_2}}}_{k_1 \cdots k_{n_2}}, ...]$;
that is, $A$ represents field components of different fields with different 
transformation properties.

We assume, as usual, a local variational principle to get the Euler--Lagrange field
equations in which $L$ denotes the Lagrangian in the form $L = L[\Phi^A,{\Phi^A}_{;k}]$, to be 
distinguished from the form $L = L^*[\Phi^A,{\Phi^A}_{,k},{\Gamma^i}_{kl}]$ :  
\begin{equation} \label{el}
\frac{\partial L^*[\Phi^C,\Phi^C_{\,\ ,l}] \sqdetR}{\partial \Phi^A}
- \frac{\partial^2}{\partial x^k}\frac{{L^*[\Phi^C,\Phi^C_{\,\ ,l}] \sqdetR}  }
{\partial{\Phi^A_{\,\ ,k}}}  = 0,
\end{equation}
which are valid for a general tensor field, $\Phi^A$, yet unspecified.  Since we want
to use covariant variables $L[\Phi^C,\Phi^C_{\,\ ;l}]$ instead of
$L^*[\Phi^C,\Phi^C_{\,\ ,l}]$, it is more appropriate to write Eq. (\ref{el}) in a
covariant form.  In order to do that, we define the covariant 
derivative through equation (\ref{covdev}).

The change from partial to covariant derivatives implies that
\begin{eqnarray}
\partq{L^*[\Phi^C,\Phi^C_{,l}]}{\Phi^A} &=& \partq{L[\Phi^C,\Phi^C_{;l}]}{\Phi^A}
+ \partq{L[\Phi^C,\Phi^C_{;l}]}{\Phi^B_{\,\ ;m}}
\partq{\Phi^B_{\,\ ;m}}{\Phi^A} \\
&=& \partq{L}{\Phi^A} + \partq{L}{\Phi^B_{\,\ ;m}}
\Gef{B}{A}{j}{i} \Gam{i}{jm}\ , \label{p1}  \\
\frac{\partial{L^*[\Phi^C,\Phi^C_{\,\ ,l}]}}{\partial{\Phi^{A}_{\,\ ,k}}} &=&
\frac{\partial{L[\Phi^C,\Phi^C_{\,\ ;l}]}}{\partial{\Phi^{A}_{\,\ ;k}}} \label{p2} 
\end{eqnarray}

Then,
\begin{equation} \label{par-cova}
\frac{\DD}{\partial x^l}\partq{L}{\Phi^A_{\,\ ;k}} =   
\frac{\partial}{\partial x^l}\partq{L}{\Phi^A_{\,\ ;k}}
+ \Gam{k}{jl}\partq{L}{\Phi^A_{\,\ ;j}}  
- \Gef{B}{A}{n}{m}\Gam{m}{nl} \partq{L}{\Phi^B_{\,\ ;k}} ,
\end{equation}
where $\frac{\DD}{\partial x^k} () \equiv ()_{;k}$, and
\be{par-cov}
\frac{\DD}{\partial x^k}\partq{L}{\Phi^A_{\,\ ;k}} =
\frac{\partial}{\partial x^k}\partq{L}{\Phi^A_{\,\ ;k}}
+ \Gam{k}{jk}\partq{L}{\Phi^A_{\,\ ;j}}  - \Gef{B}{A}{n}{m}\Gam{m}{nk} \partq{L}{\Phi^B_{\,\ ;k}}\ .
\ee
The determinant of the Ricci tensor transforms as follows
\begin{equation}\label{derdetRicci}
\partq{}{x^k}\left( \ln\sqdetR \right) =   \frac{\DD}{\partial x^k}
\left( \ln\sqdetR \right) +  \Gam{m}{mk}\ ,
\end{equation}
where we assumed that (in contrast to the notation in GR) $R^{ij}R_{jk} = \delta^{i}_{k}$.
Note that though the metric tensor also possesses this property, it is not necessarily implied a relation of the type 
given by Eq. (\ref{dS}). In fact, below we will see that in the presence of matter 
fields the Ricci and the metric tensors must be different.  Combining the above 
formulas, we obtain the tensorial equation
\begin{equation} \label{grav-eq}
\partq{L}{\Phi^A} - \frac{\DD}{\partial x^k}\partq{L}{\Phi^A_{\,\ ;k}}   
-\left[\frac{\DD}{\partial x^k} \ln\sqdetR +
2 \Gam{m}{[mk]} \right]   \partq{L}{\Phi^A_{\,\ ;k}} = 0  .
\end{equation}
This is the covariant field equation, valid for a general matter field
$\Phi^A$.

\bigskip

Up to this point, an explicit dependence of $L$ on $R_{ik}$ was not involved. 
We now turn to the more general action, Eq. (\ref{stwotwo}).
The equation for the affine connection,
\begin{equation}
 \partq{L[\Phi^C,{\Phi^C}_{\, ;l}, R_{mn}] \sqdetR}
{\Gam{a}{bc}}  
- \partq{}{x^k}\partq{L[\Phi^C,\Phi^C_{\, ;l}, R_{mn}] \sqdetR}
{\Gam{a}{bc,k}}  = 0, \label{elg}
\end{equation}
is in this form neither tensorial nor covariant. We start from the definition of the Ricci tensor,
\ben
R_{ik} \equiv \Gam{l}{il,k} - \Gam{l}{ik,l} + \Gam{l}{im}\Gam{m}{lk} - \Gam{l}{ik}\Gam{m}{lm}\ .
\een
A straightforward calculation yields
\ban
\partq{R_{ik}}{\Gam{a}{bc}} &=& \Gam{r}{st} {E_{rika}}^{stbc}\ ,\\
{E_{rika}}^{stbc}&=&
\delta^c_r\delta^b_i\delta^t_k\delta^s_a
+
\delta^b_r\delta^s_i\delta^c_k\delta^t_a
-
\delta^b_r\delta^s_i\delta^t_k\delta^c_a
-
\delta^t_r\delta^b_i\delta^c_k\delta^s_a\ ,
\\
\partq{R_{ik}}{\Gam{a}{bc,d}} &=& {D_{ika}}^{bcd} = \delta^c_a\delta^b_i\delta^d_k
- \delta^d_a\delta^b_i\delta^c_k\ .
\ean
Formally, the variational derivatives are
\ban
\frac{\delta[\sqdetR L]}{\delta[\Gam{a}{bc}]}&=&\partq{\sqdetR L}{\Gam{a}{bc}} - 
\partq{}{x^d}\partq{\sqdetR L}{\Gam{a}{bc,d}} \\
&=&
{\PP_B}^c \Gef{B}{A}{b}{a}\Phi^A 
+{E_{rika}}^{stbc} \Gam{r}{st} G^{ik} - {D_{ika}}^{bcd} \partq{}{x^d}G^{ik}\ ,
\ean
where we used the abbreviations 
\ben
\GG^{ik} \equiv \partq{\sqdetR L}{R_{ik}} \quad \mbox{and}\quad {\PP_B}^c \equiv 
\sqdetR \partq{L}{{\Phi^B}_{;c}}\ .
\een

We now solve the Euler--Lagrange equation for $G^{ik}$ through use of the relation
\ben
{D_{ika}}^{bcd} \left(\delta^a_l\delta^m_b\delta^n_c -
\frac{1}{3}\delta^a_c\delta^n_l\delta^m_b \right)
= - \delta^d_l\delta^m_i\delta^n_k
\een
and obtain

\begin{eqnarray}\label{gammaeq}
\partq{}{x^l}G^{mn} & = &
- \sqdetR \partq{L}{{\Phi^B}_{;c}} \Gef{B}{A}{b}{a}\Phi^A (\delta^a_l\delta^m_b\delta^n_c - 
\frac{1}{3}\delta^a_c\delta^n_l\delta^m_b)\\
& - & {E_{rika}}^{stbc} \Gam{r}{st} G^{ik}(\delta^a_l\delta^m_b\delta^n_c - 
\frac{1}{3}\delta^a_c\delta^n_l\delta^m_b) \nonumber\\
\weglassen{
& = &
- {\PP_B}^c \Gef{B}{A}{b}{a}\Phi^A (\delta^a_l\delta^m_b\delta^n_c - 
\frac{1}{3}\delta^a_c\delta^n_l\delta^m_b)\\ & - & \Gam{r}{st} G^{ik}(
  \delta^n_r\delta^m_i\delta^t_k\delta^s_l
+ \delta^m_r\delta^s_i\delta^n_k\delta^t_l
- \delta^t_r\delta^m_i\delta^n_k\delta^s_l
- \frac{1}{3}\delta^s_r\delta^m_i\delta^t_k\delta^n_l
+ \frac{1}{3}\delta^t_r\delta^m_i\delta^s_k\delta^n_l
) \\
}
 & = &
- {\PP_B}^n \Gef{B}{A}{m}{l}\Phi^A + \frac{1}{3}\delta^n_l{\PP_B}^s \Gef{B}{A}{m}{s}\Phi^A\\ \nonumber
& - & \Gam{n}{ls}G^{ms} - \Gam{m}{sl}G^{sn} + \Gam{s}{ls}G^{mn} + 
\frac{1}{3}\delta^n_lG^{ms}(\Gam{t}{ts}-\Gam{t}{st})
\end{eqnarray}

There are two contractions,

\ben
\partq{}{x^n}\GG^{nm} = -{\PP_B}^m\Gef{B}{A}{n}{n}\Phi^A +\frac{1}{3} 
{\PP_B}^n\Gef{B}{A}{m}{n}\Phi^A
- \Gam{m}{st}\GG^{st} + \frac{1}{3} \GG^{ms}(\Gam{t}{ts}-\Gam{t}{st})
\een

and
\ben
\partq{}{x^n}\GG^{mn} =  \frac{1}{3} {\PP_B}^n\Gef{B}{A}{m}{n}\Phi^A
- \Gam{m}{st}\GG^{st} + \frac{1}{3} \GG^{ms}(\Gam{t}{ts}-\Gam{t}{st})
\een
that is,
\ben
\partq{}{x^n}(\GG^{mn}-\GG^{nm}) = {\PP_B}^m\Gef{B}{A}{n}{n}\Phi^A\ .
\een
This is the generalization of the known relation in Schr\"odinger's theory.

In our particular case, we obtain
\begin{eqnarray} \label{grav-eq1}
&&\Gam{b}{ka} R^{ck} + \Gam{c}{ak} R^{kb} -
\Gam{k}{ak} R^{c b} - \Gam{b}{k l} R^{ l k}
\delta^{c}_{a} + 2 \frac{\partial{{\rm ln} L}}{\partial{\Gam{a}{b c}}} =
\nonumber \\
&&\left[ {\rm ln} \, L \sqdetR \right]_{, k}
\left[\delta^{c}_{a} R^{kb} - \delta^{k}_{a} R^{c b} \right] 
+ \left[\delta^{c}_{a} R^{kb} - \delta^{k}_{a} R^{c b} \right]_{, k} .
\end{eqnarray}
By contracting the indices $a$ and $c$, and substituting that equation again into
Eq. (\ref{grav-eq1}) implies that

\begin{eqnarray} \label{grav-eq2}
&&{R^{cb}}_{,a} +  \left[ {\rm ln}\sqdetR \right]_{, a} R^{c b} + 
\left[ \frac{2}{3} \Gam{l}{[kl]} \delta^{c}_{a}
- \Gam{l}{al} \delta^{c}_{k} +  \Gam{c}{ak}
\right] R^{kb} + \Gam{b}{ka} R^{ck} = \nonumber \\ && 
 -  2 \left[
\frac{\partial{{\rm ln} L}}{\partial{\Gam{a}{bc}}} -
\frac{1}{3} \frac{\partial{{\rm ln} L}}{\partial{\Gam{r}{bs}}}
\delta^{r}_{s}\delta^{c}_{a} \right]  \, , 
\end{eqnarray}
where the term $\frac{\partial{{\rm ln} L}}{\partial{\Gam{a}{b c}}}$ can be further given as
\be{transg}
\frac{\partial{{\rm ln} L}}{\partial{\Gam{a}{b c}}} =
\frac{\partial{{\rm ln} L}}{\partial{\Phi^A_{\,\ ; c}}} \Gef{A}{B}{b}{a}\Phi^B  \, .
\ee

Schr\"odinger discovered that by defining a new affinity,
${}^*\Gam{a}{bc} \equiv \Gam{a}{bc} + \frac{2}{3} \delta^{a}_{b}\Gam{l}{cl}$, 
equation (\ref{grav-eq2}) with $L=$const. reduces to 
$R^{c b}_{\,\  {}^{\ast}_{,}  a } \equiv 
R^{c b}_{\,\  ,a} + {}^*\Gam{c}{ka} R^{kb} + {}^*\Gam{b}{ak} R^{ck} = 0$. 
In our case, these definitions imply that 

\begin{eqnarray}\label{dercovmatt}
R^{c b}_{\,\  {}^{\ast}_{,}  a } &=&
- 2 \left[\frac{\partial{{\rm ln} L}}{\partial{{}^*\Gam{a}{bc}}} -
\frac{1}{3}\left[ \frac{\partial{{\rm ln} L}}{\partial{{}^*\Gam{k}{kc}}}
\delta^{b}_{a} +  \frac{\partial{{\rm ln} L}}{\partial{{}^*\Gam{k}{bk}}}
\delta^{c}_{a} \right] \right]   
\nonumber \\ 
&+&
\left[\frac{\partial{{\rm ln} L}}{\partial{{}^*\Gam{a}{kl}}} -
\frac{1}{3}\left[\frac{\partial{{\rm ln} L}}{\partial{{}^*\Gam{k}{kl}}}
\delta^{m}_{a} +
\frac{\partial{{\rm ln} L}}{\partial{{}^*\Gam{k}{mk}}} \delta^{l}_{a}
\right] \right] R_{ml} R^{c b} .
\end{eqnarray}
The introduction of matter fields ($L\neq$const.) avoids $R^{c b}$ being parallel 
transported into itself by the star affinity; the same holds for the Einstein 
affinity, see Ref. \cite{THJ94}.  Then, the presence of matter fields
preclude us to interprete the Ricci tensor as being the metric, see Eq. (\ref{dS}). 

\section{The metric of space--time in the shock-wave picture} \label{secIV}

We identify the metric through
the propagation of shock waves. The observation of the propagation of (shock) waves
defines the metric of the wave in question. In ordinary wave mechanics, the wave operator
determines the shocks to propagate along its bisectrices. Each wave equation 
has its own causal cone when the wave operators differ in the highest order of derivatives.
The principle of relativity requires that the
propagation is the same for the different fields that one intends to include as fundamental,
but this is a second question. In a construction like the action given by 
Eq. (\ref{stwoone}),
the propagation of shock waves is given through substitution of
\ben
\Phi_{\rm shock} = \Phi_0 + \theta[z] z^2 \phi
\een
for the fields $\Phi$, where $z = z[x^k] = 0$ defines the shock
hypersurface, $\Phi_0$ and $\phi$ are at least $C_2$ in a
neighborhood of the shock. The difference in the second-order
derivatives of the two sides of the hypersurface is
\ben
\Delta(\Phi_{,ik}) = \phi~z_{,i}z_{,k}
\een
On the shock front, the Euler-Lagrange equation requires \cite{BLD95}:
\ben
\frac{\partial^2 L}{\partial \Phi^A_{,i}\partial \Phi^B_{,k}} {\phi}^B z_{,i}z_{,k} = 0\ .
\een
In the case of only one scalar field, the result is trivially
\ben
g^{ik} \propto \frac{\partial^2 L}{\partial \Phi_{,i}\partial \Phi_{,k}}
\een

In the case of more than one field component,
we obtain a component-dependent propagation of the form
\ben
{K_{AB}}^{ik}{\phi}^B~z_{,i}z_{,k} = 0 \, ,
\een
where ${K_{AB}}^{ik} \equiv \frac{\partial^2 L}{\partial \Phi^A_{,i}\partial \Phi^B_{,k}}$. 
Local Lorentz invariance requires that the light-cones at least for the fundamental
free fields coincide. Therefore, GR implies the separability
\ben
{K_{AB}}^{ik}  =  a_{AB} g^{ik}
\een
in order to obtain equal propagation cones for all field components \cite{ABL93,BLD95}.

Note that the coefficients $K_{AB}{}^{ik}$  depend on the construction of $L$, and not on the
volume element $\sqdetR\dd^4x$.
The space--time Ricci curvature is irrelevant for the propagation of the shocks as long as
it is not \textit{explicitly} used in forming 
$L_{\rm matter}[\Phi^A, {\Phi^A}_{;k}, R_{ik}]$. However,
explicit use implies higher order non-linearity, again.
 
\section{Local action integrals} \label{secV}

Let us assume a contravariant vector field $\Phi^k$. When the Ricci tensor enters 
the action through the volume element only, we can construct actions such as 
\be{sthree}
S_3 = \int (\alpha\ \Phi^k_{\, ;l} \Phi^l_{\, ;k}+ \beta\ \Phi^k_{\, ;k} \Phi^l_{\, ;l})
\sqdetR\,  \dd^4x
\ee
and we obtain
\ben
{K_{ab}}^{ik} = \frac{\partial^2 L}{\partial {\Phi^a}_{;i}\partial {\Phi^b}_{;k}}  
\propto 2\ \alpha\ \delta^k_a\delta^i_b
+ 2\ \beta\  \delta^i_a\delta^k_b
\een
and, therefore for arbitrary $\alpha$ and $\beta$ such that $\alpha + \beta \neq 0$,  one has that
\ben
\phi^b z_{,b} = 0.
\een
This is a limitation only for the amplitude of the shock, and no limitation for its front.
Again, the form of the volume element does not enter the shock condition. In a local 
theory, its construction cannot yield the metric of space--time.  

It is not  difficult to 
see that in a local theory any propagation depends on the local amplitudes of the  interacting 
fields and not on the geometry of the shock fronts as long as the lightcones are not 
deliberately constructed through use of some second--order contravariant tensor field, i.e. an 
a priori metric. Such an a priori metric however destroys our program, and cannot be its solution.
\bigskip

Let us now take Schr\"odinger's choice to go around the fatal result that the matter fields 
itself cannot locally determine a viable light cone. We still stick to a local construction and 
replace the ordinary metric with the Ricci tensor. This might not be the final construction because 
one expects, at least approximately, metricity of the connection \cite{BHH97}, but we need 
only the shock approximation and the qualitative features of the field equations for our 
argument. We consider an action of the type given by Eq. (\ref{stwotwo}) 
that is constructed using the methods of GR or of the metric-affine 
theory followed by a substitution of $R_{ik}$ for $g_{ik}$ (the undifferentiated $g_{ik}$, not 
in the connection $\Gam{a}{bc}$). The propagation of matter fields, of course, follows now the 
cone that is determined by $R^{ik}$ as constructed.  However, the field equation for the 
connection now yields a restricting condition for the decisive part of the energy-momentum 
tensor density,

\ben
\TT^{ik} = \frac{\delta[L[g_{..},\Phi^A,({\Phi^A}_{,m}+ 
\Gef{A}{B}{b}{a}\Gam{a}{bm}[g..,g..,.]\Phi^B)]\sqdetg]}{\delta[g_{ik}]}\ ,
\een

namely,
\ben
\GG^{ik} = \partq{(L[g_{..},\Phi^A,{\Phi^A}_{;m}]\sqdetg)}{g_{ik}}|_{{\rm at\ } g..=R..}\ ,
\een
where the implicit dependence of ${\Phi^A}_{;m}$ on $g_{ik}$ and its derivatives does not enter.
We arrive at field equations for the connection, Eq. (\ref{gammaeq}),
that restrict the energy-momentum tensor density to kind of
constant values, i.e., to peculiar, and not general, physical cases.

\bigskip

Summarizing: Local theories of the type given by Eqs. (\ref{stwoone}) and (\ref{stwotwo})   
do not achieve a viable causal structure. In the former case, when the Ricci tensor only enters
the volume element, 
the shock waves of the matter fields do not 
feel that metric and are not null surfaces as expected.  In the latter case, when 
the Ricci tensor is deliberately substituted for the metric, the Schr\"odinger 
result of a covariantly constant Ricci tensor 
turns into a correspondingly constant matter tensor and excludes nearly all physical 
cases. It was our intention to show this in due generality.  

\bigskip

We may use matter fields to construct a volume element in order to get field equations
in space--times without curvature, too \cite{ATH83,GMM98}. In doing that, it is difficult not to
introduce an a priori metric. Akama and Terazawa \cite{ATH83} hide it in the summation of 
their scalar fields, Gronwold et al \cite{GMM98} have it explicitly in their 
Lagrangians (see their section IV).

The construction of a metric through local non-gravitational fields has the
consequence that a strong dependence of the metric on local perturbation must be
expected. For instance, the metric components should be expected to be
proportional to the local mass already at zeroth order. Therefore, we conclude that:

1. an a posteriori observation-based definition of a metric must rely on
non-gravitational fields even in presence of a curved affine
connection, and

2. its definition requires an explicitly non--local action for
the non--gravi\-ta\-tional fields.

\section{Non-local action integrals and final remarks} \label{secVI}

The concluding remark shall discuss non-local action integrals.
When we construct action integrals with fields and connections alone
we find field equations that exist only in the case when both $\Phi^A$ and $R_{ik}$
are non-trivial. If there is no matter, the geometry cannot be measured and is
free. If there is no curvature, the motion of matter is not defined.
It is, of course, a drawback in the local action that matter has to exist 
locally in order to have a geometry defined. We think that matter 
somehow should be enough to fix a geometry, like in Machian 
approaches. A non--local Lagrangian will be the next step.

A non-local interaction is constructed through at least a twofold integration over
space--time, such as 
\ban
S_5 &=& \int \dd^4x \int \dd^4y \sqrt{\detR[x]} \sqrt{\detR[y]} \\
& \times &L\left[ \Phi^A[x], \Phi^A_{,i}[x], R_{ij}[x], \Phi^B[y], \Phi^B_{,k}[y], R_{kl}[y] \right]
\ean
where, for instance, the $x-$coordinate can be used to label a local integration and the 
$y-$coordinate to refer the rest of the world.  The non--local interaction, however,  is a delicate 
point to be constructed properly.  As long as the Lagrangian $L$ can be expanded in a series of 
scalar functions at $x$ with coefficients that are scalar functions at $y$, for instance,
\begin{equation}
S_5 = \int \dd^4x \int \dd^4y \sqrt{\detR[x]} \sqrt{\detR[y]} \nonumber \\
  \,  \sum\limits_\alpha X_\alpha[x]Y_\alpha[y]\ ,
\end{equation}
one of the integrations can formally performed. The result is a local action,
\ben
S_5 = \int \dd^4x \sqrt{\detR[x]} \sum\limits_\alpha \eta_\alpha X_\alpha[x]
\een
with coefficients
\ben
\eta_\alpha = \int \dd^4y \sqrt{\detR[y]} Y_\alpha[y]\ .
\een
No new physics is found.

When we now try to implement terms like $\Psi^k[x]\Phi_k[y]$, we have to see 
that they are not scalars at all: $\Psi^k[x]$ is a vector for substitutions 
of $x$, and a scalar for substitutions of $y$.  On the opposite, $\Phi_k[y]$ is 
a scalar for substitutions of $x$, and a vector for substitutions of $y$.  The product 
can be made a scalar for both substitutions only when there exist a bi-tensor
${\gamma_l}^m[x,y]$ that depends on the two points, and transforms as a covariant 
vector when the $x$ are substituted, and as a contravariant vector when $y$ is 
substituted \cite{SJL60}.  In this case, $\Psi^l[x]{\gamma_l}^m[x,y]\Phi_m[y]$ is a 
scalar and may be used in constructing Lagrangians.

With the bi--tensor $\gamma$, however, we introduced an additional 
teleparallel connection \cite{SJL60}:
\ben
{}^{**}\Gam{a}{bc} = - \partq{}{y^c}{\gamma_b}^a[x,y]\mid_{{\rm at}\ y=x}
\een
This makes the connection $\Gam{a}{bc}$ superfluous.
In addition, the connection $\Gam{a}{bc}$ is now equivalent to a tensor
field, ${T^a}_{bc} = \Gam{a}{bc} - {}^{**}\Gam{a}{bc}$, of third order.
Using  ${}^{**}\Gam{a}{bc}$, we lose the Ricci tensor as equivalent of some metric:
The curvature
of a teleparallel connection vanishes, and the Schr\"odinger choice must be
replaced by some other construction.
In addition, it is important to note that the bi-tensor $\gamma$ is mixed-variant.
A covariant bi-tensor $\gamma_{ik}[x,y]$ is the generalization of the metric tensor 
$g_{ik}[x] = \lim\limits_{y\rightarrow x} \gamma_{ik}[x,y]$. Teleparallel theories 
are discussed in connection with e.g. string theory
or with rotation in the universe \cite{OCS03,OPJ03}; but these are different approaches 
that are out of the scope of the present work.



\def\GRG{Gen. Rel. Grav.}
\def\PR{Phys. Rev. }
\def\PL{Phys. Lett. }
\def\AN{Astron. Nachr. }
\def\AdP{Ann. d. Phys.(Lpz.) }
\def\PRSIrA{Proc. Roy. Irish Acad. }
\newcommand{\bb}[8]{
\bibitem{#2}{#1}, {#7} 
 {\it #4} {\bf #5}, #6 (#3).}

\end{document}